\documentclass[twocolumn,prl,showpacs,amsmath,amssymb]{revtex4}
\usepackage[latin1]{inputenc}
\usepackage{graphicx}
\usepackage{dcolumn}
\usepackage{bm}

\begin{document}



\title{Quantum interference oscillations of the superparamagnetic blocking in an Fe$_{8}$ molecular nanomagnet.}

\author{E. Burzur\'{\i}}
\affiliation{Instituto de Ciencia de Materiales de Arag\'on (ICMA), C.S.I.C. - Universidad de Zaragoza, E-50009 Zaragoza, Spain}

\author{F. Luis}
\email{fluis@unizar.es} \affiliation{Instituto de Ciencia de Materiales de Arag\'on (ICMA), C.S.I.C. - Universidad de Zaragoza, E-50009 Zaragoza, Spain}

\author{O. Montero}
\affiliation{Instituto de Ciencia de Materiales de Arag\'on (ICMA), C.S.I.C. - Universidad de Zaragoza, E-50009 Zaragoza, Spain}

\author{B. Barbara}
\affiliation{Institut N\'eel, CNRS \& Universit\'e Joseph Fourier, BP166, 38042 Grenoble Cedex 9, France}

\author{R. Ballou}
\affiliation{Institut N\'eel, CNRS \& Universit\'e Joseph Fourier, BP166, 38042 Grenoble Cedex 9, France}

\author{S. Maegawa}
\affiliation{Graduate School of Human and Environmental Studies, Kyoto University, Kyoto 606-8501, Japan}


\date{\today}


\begin{abstract}
We show that the dynamic magnetic susceptibility and the superparamagnetic blocking temperature of an Fe$_{8}$ single molecule magnet oscillate as a function of the magnetic field $H_{x}$ applied along its hard magnetic axis. These oscillations are associated with quantum interferences, tuned by $H_{x}$, between different spin tunneling paths linking two excited magnetic states. The oscillation period is determined by the quantum mixing between the ground $S=10$ and excited multiplets. These experiments enable us to quantify such mixing. We find that the weight of excited multiplets in the magnetic ground state of Fe$_{8}$ amounts to approximately $11.6$ \%.
\end{abstract}

\pacs{}

\maketitle



\begin{figure}[h]
\resizebox{7.5cm}{!}{\includegraphics{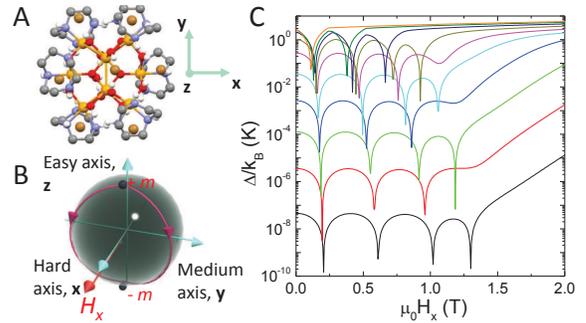}}
\caption{A: Molecular structure of the [(C$_{6}$H$_{15}$N$_{3}$)$_{6}$Fe$_{8}$O$_{2}$(OH)$_{12}$] molecular magnet, briefly referred to as Fe$_{8}$. B: Two equivalent tunneling paths linking states with $m=+S$ and $m=-S$. A magnetic field along the hard axis (denoted by $x$) shifts the relative phases of these trajectories, thus leading to constructive and destructive interferences. C: Dependence of the quantum tunnel splittings $\Delta_{m}$ on $H_{x}$ calculated with Eq. (\ref{hamilt1}) for states $\pm m$, with $m = 10$ (bottom curve) to $m =1$ top curve).}
\label{deltaVsH}
\end{figure}

High-spin molecular clusters \cite{Christou2000,Gatteschi2003} display superparamagnetic behavior, very much as magnetic nanoparticles typically do. Below a time- (or frequency-)dependent blocking temperature $T_{\rm b}$, the linear magnetic response "freezes" \cite{Hernandez1996,Barra1996} and magnetization shows hysteresis \cite{Sessoli1993}. The slow magnetic relaxation of these single-molecule magnets (SMMs) arises from anisotropy energy barriers separating spin-up and spin-down states. Because of their small size, the magnetic response shows also evidences for quantum phenomena, such as resonant spin tunneling \cite{Hernandez1996,Friedman1996,Thomas1996,Sangregorio1997}. In the case of molecules that, like Fe$_{8}$ (cf Fig. \ref{deltaVsH}A and \cite{Barra1996}), have a biaxial magnetic anisotropy, tunneling between any pair of quasi-degenerate spin states $\pm m$ can proceed via two equivalent trajectories, which, as illustrated in Fig. \ref{deltaVsH}B, cross the hard anisotropy plane close to the medium anisotropy axis. A magnetic field along the hard axis changes the phases of these tunneling paths, leading to either constructive or destructive interferences. This phenomenon is known as Berry phase interference \cite{Berry1984,Garg1993}.

Experimental evidences for the ensuing oscillation of the quantum tunnel splitting $\Delta_{m}$, shown in Fig. \ref{deltaVsH}C, were first observed in Fe$_{8}$ \cite{Wernsdorfer1999,Wernsdorfer2000} and then in some other SMMs \cite{Wernsdorfer2002,Lecren2005,Wernsdorfer2005,Ramsey2008,Wernsdorfer2008,DelBarco2010,Quddusi2011} by means of Landau-Zener magnetization relaxation experiments. Interference patterns measured on Fe$_{8}$ at very low temperatures, which correspond to tunneling via the ground state doublet $m = \pm 10$, are reproduced by the following spin Hamiltonian

\begin{equation}
{\cal H} = -DS^{2}_{z} + E(S^{2}_{x} - S^{2}_{y}) + C \left( S_{+}^{4} + S_{-}^{4} \right)  - g\mu_{\rm{B}}\overrightarrow{S}\cdot\overrightarrow{H}
\label{hamilt1}
\end{equation}

\noindent that applies to the lowest lying spin multiplet, with $S=10$, and where $D/k_{\rm B} = 0.294$ K, $E/k_{\rm B} = 0.046$ K, $C/k_{\rm B} = -2.9 \times 10^{-5}$ K are magnetic anisotropy parameters, and $g=2$. The sizeable fourth-order parameter $C$ reflects not only the intrinsic anisotropy but, mainly, it parameterizes quantum mixing of the $S=10$ with excited multiplets ($S$-mixing) and how it influences quantum tunneling via the ground state \cite{Carretta2004}.

In the present paper, we study the influence of Berry phase interference on the ac magnetic susceptibility $\chi$ and $T_{\rm b}$ of Fe$_{8}$, that is, on those quantities that characterize the standard SMM (or superparamagnetic) behavior. Close to $T_{\rm b}$, magnetic relaxation is dominated by tunneling near the top of the anisotropy energy barrier, thus also near excited multiplets with $S \neq 10$. In this way, we aim also to use the interference pattern to gain quantitative information on the degree of $S$-mixing in Fe$_{8}$.


The sample employed in these experiments was a $3 \times 2 \times 1$ mm$^{3}$ single crystal of Fe$_{8}$. Each molecule has a net spin $S = 10$ and a strong uniaxial magnetic anisotropy. Equation (\ref{hamilt1}) defines $x$, $y$ and $z$ as the hard, medium and easy magnetization axes. In the triclinic structure of Fe$_{8}$, $x$, $y$, and $z$ axes are common to all molecules \cite{Ueda2001}. The complex magnetic susceptibility $\chi = \chi^{\prime}(T,\omega)-i\chi^{\prime \prime}(T,\omega)$ was measured between $90$ mK and $7$ K, and in the frequency range $3$ Hz $\leq \omega/2\pi \leq 20$ kHz, using a purpose built ac susceptometer thermally anchored to the mixing chamber of a $^{3}$He-$^{4}$He dilution refrigerator. A dc magnetic field $\overrightarrow{H}$ was applied with a $9$ T$\times1$ T$\times 1$ T superconducting vector magnet, which enables rotating $\overrightarrow{H}$ with an accuracy better than $0.001^{\circ}$. The magnetic easy axis $z$ was parallel to the ac excitation magnetic field of amplitude $h_{ac} = 0.01$ Oe. The sample was completely covered by a non-magnetic epoxy to prevent it from moving under the action of the applied magnetic field. The alignment of $\overrightarrow{H}$ perpendicular ($\pm 0.05^{\circ}$) to $z$ and close ($\pm 5^{\circ}$) to the hard $x$ axis was done at low temperatures ($T = 2$ K), using the strong dependence of $\chi^{\prime}(T,\omega)$ on the magnetic field orientation (see \cite{Burzuri2011} for further details). These data were scaled with measurements performed, for $T \geqslant 1.8$ K, using a commercial SQUID magnetometer and a physical measurement platform equipped with ac susceptibility options.


\begin{figure}[h]
\resizebox{7.5cm}{!}{\includegraphics{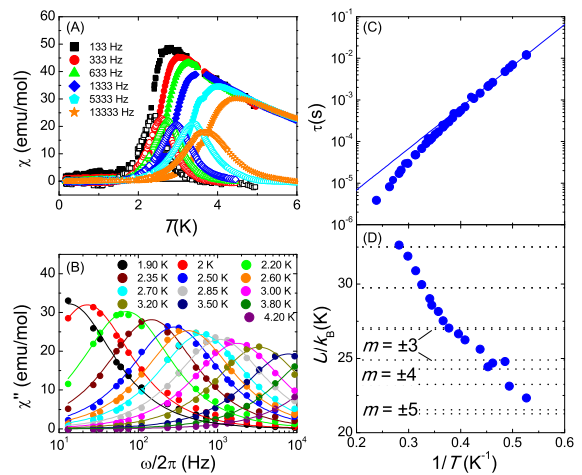}}
\caption{Ac susceptibility of an Fe$_{8}$ single crystal measured at several frequencies. A: temperature dependence of the real (solid symbols) and imaginary (open symbols) components. B: frequency dependence of the imaginary component. Lines are least-square fits with Cole-Cole function $\chi^{\prime \prime} (\omega , T) = \Delta \chi(\omega \tau)^{\beta} \sin(\beta \pi/2)/[1+(\omega \tau)^{2 \beta}+2(\omega \tau)^{\beta} \cos(\beta \pi/2)]$, where $\Delta \chi \simeq \chi_{T}$, the equilibrium susceptibility, and $\beta \simeq 0.92$. C: Arrhenius plot of the relaxation time $\tau$ extracted from these fits. The solid line is a least-squares linear fit of data measured below $2.6$ K. D: Effective activation energy $U$ for the magnetic relaxation process, obtained from the slope of the Arrhenius plot. The horizontal lines show magnetic energy levels derived from the giant spin Hamiltonian (\ref{hamilt1}).}
\label{XvsT}
\end{figure}

The zero field ($H=0$) ac susceptibility $\chi^{\prime}$ and $\chi^{\prime \prime}$ components of Fe$_{8}$ show the typical SMM behavior (Fig. \ref{XvsT}). Maxima of $\chi^{\prime \prime}$ measured at different frequencies define $T_{\rm{b}}$ and occur when $\tau \simeq 1/\omega$, where $\tau$ is the magnetic relaxation time. As Fig. \ref{XvsT} shows, $\tau$ approximately follows Arrhenius' law $\tau  \simeq \tau_{0} \exp \left(U/k_{\rm B}T \right)$, where $U$ is the activation energy and $\tau_{0}$ is an attempt time. However, a closer inspection reveals that the slope of the Arrhenius plot increases gradually with temperature, from a low-$T$ value $U/k_{\rm B} \simeq  22$ K to more than $32$ K. While the former $U$ value agrees with tunneling taking place via $m = \pm 5$ states, the latter is close to the maximum energy ($\simeq 32.5$ K) of the $S=10$ multiplet.

\begin{figure}[h]
\resizebox{7.5cm}{!}{\includegraphics{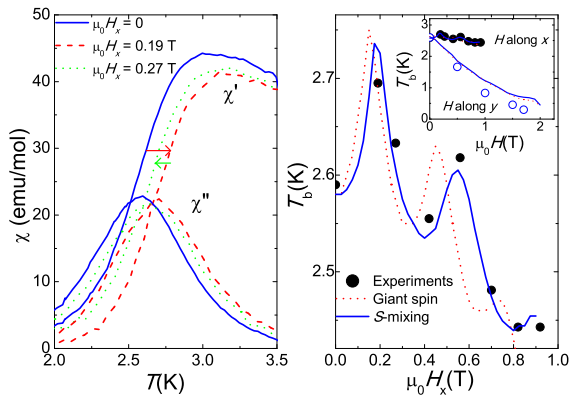}}
\caption{Left: $\chi^{\prime}$ and $\chi^{\prime \prime}$ susceptibility components of Fe$_{8}$ measured at $\omega/2\pi=333$ Hz and for three different $H_{x}$ values. Right: blocking temperature $T_{\rm b}$ as a function of $H_{x}$ (hard axis) and $H_{y}$ (medium axis) for the same frequency. Dotted and solid lines are theoretical predictions following from, respectively, the giant spin model [Eq. (\ref{hamilt1})] and the two-spin model [Eq. (\ref{hamilt2})], which includes $S$-mixing effects, for $\phi=4$ deg. The inset compares blocking temperatures measured with $\vec{H}$ along the hard ($x$) and medium ($y$) axes.}
\label{XvsTHn0}
\end{figure}

Figure \ref{XvsTHn0} shows $\chi^{\prime}(T)$ and $\chi^{\prime \prime}(T)$ data measured at $\omega/2\pi$ = 333 Hz and under three different transverse magnetic fields. By increasing $\mu_{0}H_{x}$ from $0$ to $0.19$ T, the superparamagnetic blocking shifts towards higher $T$. Thereafter, $T_{\rm{b}}$ decreases again with further increasing $\mu_{0}H_{x}$ to $0.27$ T. The right-hand panel of Fig. \ref{XvsTHn0} shows that $T_{\rm{b}}$ oscillates as a function of $H_{x}$. This behavior contrasts sharply with the rapid and monotonic decrease of $T_{\rm{b}}$ that is observed when $\overrightarrow{H}$ is parallel to the medium axis $y$ (see the inset of Fig. \ref{XvsTHn0} and \cite{Burzuri2011,Luis2000}).

\begin{figure}[h]
\resizebox{7.5cm}{!}{\includegraphics{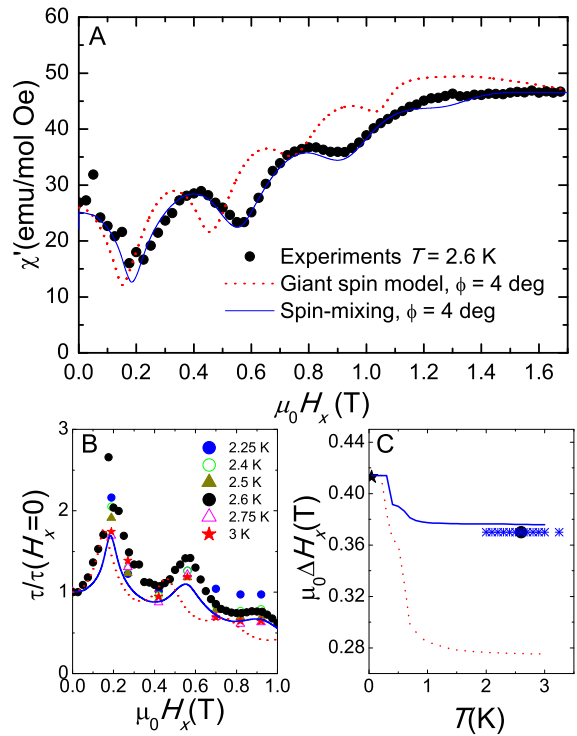}}
\caption{A: $\chi^{\prime}$ of Fe$_{8}$ {\em vs} $H_{x}$ measured at $T = 2.6$ K and $\omega/2\pi$ = 333 Hz. B: transverse field dependence of the magnetic relaxation time $\tau$ determined at different temperatures. C: temperature dependence of the quantum oscillation periods estimated from Landau-Zener relaxation measurements ($\star$) \cite{Wernsdorfer1999}, $\chi^{\prime}$ vs $H_{x}$ at $T=2.6$ K ($\bullet$), and $\chi^{\prime}$ vs $T$ at different $H_{x}$ ($\ast$). Dotted and solid lines are theoretical predictions, for $\phi=4$ deg., that follow from the giant spin model [Eq. (\ref{hamilt1})] and the two-spin model [Eq. (\ref{hamilt2})], respectively.}
\label{XvsH}
\end{figure}

Oscillations of $T_{\rm b}$ lead also to oscillations of the dynamical susceptibility. Figure \ref{XvsH}A shows $\chi^{\prime}$ vs $H_{x}$ data measured at $\omega/2\pi = 333$ Hz and $T = 2.6 \simeq T_{\rm b}(H_{x} = 0)$ K. Under such conditions, small shifts of $T_{\rm{b}}$ result in large changes of $\chi^{\prime}$, thus allowing us to monitor these changes very precisely. $\chi^{\prime}$ shows three minima, at $\mu_{0}H_{\rm a} \simeq 0.20(1)$ T, $\mu_{0}H_{\rm b} \simeq 0.56(1)$ T, and $\mu_{0}H_{\rm c} \simeq 0.90(1)$  T, with an approximate periodicity $\mu_{0} \Delta H_{x} \equiv \mu_{0} \left[  2H_{\rm a} + (H_{\rm b}-H_{\rm a}) + (H_{\rm c}-H_{\rm b}) \right]/3 \simeq 0.37$ T. Again, this behavior contrasts with the abrupt increase towards equilibrium that is observed when $\vec{H}$ is applied along $y$ \cite{Burzuri2011}. From the susceptibility we estimate also $\tau \simeq [r/(\sin \beta \pi/2 - r \cos \beta \pi/2)]^{(1/\beta)}/\omega$, where $r= \chi^{\prime \prime}/\chi^{\prime}$ and $\beta \simeq 0.92$ was determined from Cole-Cole fits performed at $H_{x} = 0$ (see Fig. \ref{XvsT}B). Figure \ref{XvsH}B shows that, as one could anticipate, $\tau$ also oscillates with $H_{x}$. Data of Figs. \ref{XvsTHn0} and \ref{XvsH} strongly suggest that the oscillation period $\Delta H_{x}$ remains approximately constant in the temperature range between $2$ and $3$ K covered by present experiments.

The oscillations can be {\em qualitatively} accounted for by recalling the dependence of the quantum tunnel splittings $\Delta_{m}$ on $H_{x}$ (Fig. \ref{deltaVsH}). At zero field and close to $T=2.6$ K, magnetic relaxation takes place mainly via thermally activated $m = \pm 4$ states. By increasing $H_{x}$, $\Delta_{4}$ gets periodically quenched and therefore tunneling is inhibited. This leads to an increase of $\tau$ and thus also of $T_{\rm b}$, as it is observed experimentally.

However, the giant spin model Eq.(\ref{hamilt1}) is unable to provide a {\em quantitative} description of the interference pattern. The magnetic relaxation time and the frequency dependent-susceptibility have been calculated by solving a Pauli master equation for the populations of all energy levels of (\ref{hamilt1}), following the model described in \cite{Luis1998}. For simplicity, we have simulated the effect that dipolar interactions between different Fe$_{8}$ clusters have on the spin tunneling probabilities \cite{Luis1998,Prokof'ev1996} by introducing an effective bias field $\mu_{0}H_{\rm{d}z} = 31$ mT. This value is the width of the distribution of dipolar fields in a magnetically unpolarized Fe$_{8}$ crystal \cite{Wernsdorfer1999b,Fernandez2005}. The results are compared to experimental data in Figs. \ref{XvsTHn0} and \ref{XvsH}. Although they show oscillations, the theoretical period $\mu_{0} \Delta H_{x} = 0.28$ T is about $20$ \% smaller than the experimental one. It is important to emphasize that the discrepancy cannot be ascribed to a small uncertainty in angle $\phi$. All theoretical curves with distinct oscillations (for $\phi \lesssim 10$ deg.) show $\Delta H_{x}$ smaller than observed.

The discrepancy originates instead from the fact that $\mu_{0} \Delta H_{x}$ obtained from Eq. (\ref{hamilt1}) strongly decreases with $m$, from $0.4$ T for $m = \pm 10$ to $0.25$ T for $m = \pm 2$. This dependence arises from the presence of a strong fourth-order anisotropy term. The same effect occurs for higher-order terms. The giant spin approximation, which neglects all excited multiplets, is therefore unable to simultaneously account for the quantum interference patterns observed at low and high-$T$.

\begin{figure}[h]
\resizebox{6cm}{!}{\includegraphics{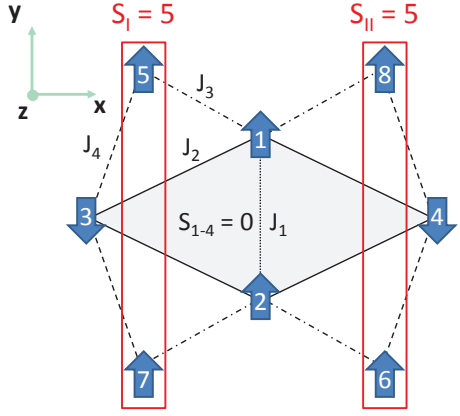}}
\caption{Scheme of exchange pathways connecting Fe$^{3+}$ ions in the Fe$_{8}$ core. Approximate values of the exchange constants are \cite{Barra2000} $J_{1}/k_{\rm B} = -36$ K, $J_{2}/k_{\rm B} = -201$ K, $J_{3}/k_{\rm B} = -26$ K, and $J_{4}/k_{\rm B} = -59$ K.}
\label{S-mixing}
\end{figure}

These results call for a more complete description, able to explicitly incorporate the effects of $S$-mixing. A schematic diagram of the magnetic structure of the Fe$_{8}$ molecular core \cite{Delfs1993,Barra2000} is shown in Fig. \ref{S-mixing}. The central diamond (or "butterfly") of spins $1-4$ strongly couple antiferromagnetically to give a net spin $S_{1-4} \simeq 0$. Couplings via the butterfly generate effective interactions between the remaining spins $5-8$. The resulting spin configuration, with a ground state $S=10$ and a first excited $S=9$ multiplet lying about $\delta E_{9,10}/k_{\rm B} = 44$ K above it \cite{Delfs1993,Barra2000,Caciuffo1998,Carretta2006}, follows from the fact that $|J_{4}| > |J_{3}|$ (all couplings are antiferromagnetic). In addition to symmetric exchange interactions, one has to consider also single-ion magnetic anisotropies, dipolar interactions and antisymmetric Dzyaloshinskii-Moriya (DM) interactions, which mix states of different $S$ \cite{Carretta2004}. Concerning the latter, although it is in principle possible to make DM interactions irrelevant on a specific bond by a gauge transformation, compatibility conditions must be satisfied to extend this over the molecule. To be precise, the product of the gauge transformations associated to each bond along any closed exchange path in the molecule should be equal to $(-1)^n I$ ($I \equiv$ identity, $n$ integer) \cite{Shekhtman1992}. The low symmetry of Fe$_{8}$ ensures that no such global similarity transformation should exist, mainly because the DM interactions between the individual spins are certainly non uniform.

Based on the above considerations, we reduce the full spin Hamiltonian of $8$ Fe$^{3+}$ ions to a simpler and computationally more affordable one, involving only two spins $S_{\rm I} =5$ and $S_{\rm II} = 5$, defined in Fig. \ref{S-mixing}. This approximation can be justified by the fact that exchange interactions in Fe$_{8}$ as well as in other SMMs have been treated by iteratively coupling spins in pairs \cite{Delfs1993,Anfuso2004}. The two-spin Hamiltonian reads as follows

\begin{eqnarray}
{\cal H} = & & \sum_{\rm i=I,II} {\cal H}_{\rm anis,i} - \sum_{\rm i=I,II} g_{\rm i} \mu_{\rm{B}}\overrightarrow{S_{\rm i}}\cdot\overrightarrow{H} \nonumber \\
 &-&  J \overrightarrow{S_{\rm I}}\overrightarrow{S_{\rm II}} - \overrightarrow{S_{\rm I}} \hat{A} \overrightarrow{S_{\rm II}}  +  \overrightarrow{d_{\rm I,II}} \overrightarrow{S_{\rm I}} \times \overrightarrow{S_{\rm II}}
\label{hamilt2}
\end{eqnarray}

\noindent where ${\cal H}_{\rm anis,i} = -D_{\rm i}S_{{\rm i},z}^{2} + E_{\rm i}(S_{{\rm i},x}^{2}- S_{{\rm i},y}^{2}) + C_{\rm i} \left( S_{{\rm i},+}^{4} + S_{{\rm i},-}^{4} \right)$ accounts for the magnetic anisotropy of each spin, $J > 0$ is an effective isotropic exchange constant, $\hat{A}$ is an anisotropic coupling tensor, and $\overrightarrow{d_{\rm I,II}}$ is a DM interaction vector. We set $D_{\rm I}/k_{\rm B} = D_{\rm II}/k_{\rm B} = 0.625$ K, $E_{\rm I}/k_{\rm B} = E_{\rm II}/k_{\rm B} = 8.94 \times 10^{-2}$ K and $C_{\rm I}/k_{\rm B} = C_{\rm II}/k_{\rm B} = -5.7 \times 10^{-5}$ K, which give, for the $S = 10$ multiplet, parameters $D = 0.47 D_{\rm I}$, $E = 0.47 E_{\rm I}$ and $C = 0.128 C_{\rm I}$ that agree with those determined from EPR experiments \cite{Barra2000}.

Within this model, dominant symmetric exchange interactions (mainly $J_{2}$ and $J_{4}$) contribute to the formation of the two giant spins $\vec{S_{\rm I}}$ and $\vec{S_{\rm{II}}}$, which are then coupled by a weaker effective symmetric exchange. We have set $J/k_{\rm B} = 3.52$ K to fit the energy gap $\delta E_{9,10}$ between the $S=9$ and the $S=10$ multiplets. The last two terms in Eq. (\ref{hamilt2}) parameterize the effects that dipolar and DM interactions, considered as perturbations, have on the energies of the subspace defined by different $\vec{S_{\rm I}}$ and $\vec{S_{\rm II}}$ orientations. Dipolar interactions between spins $(5,7)$ and $(6,8)$ give predominantly rise to a term $-A_{xx}S_{{\rm I},x}S_{{\rm II},x}$, with $A_{xx}/k_{\rm B} \simeq 2.8 \times 10^{-2}$ K. This term hardly has any noticeable influence on the period of quantum oscillations and, furthermore, it tends to {\em reduce} $\Delta H_{x}$. Because of the close to planar molecular structure and its pseudo $C_{2}$ symmetry, the same considerations apply to terms arising from dipolar interactions between any of these spins and those forming the central butterfly. In order to account for the observations, $S$-mixing must therefore predominantly arise from antisymmetric exchange interactions.

DM interactions between individual Fe$^{3+}$ spins are generally of the order of $\Delta g/g \approx 0.01$ times the symmetric interactions \cite{Moriya1960,Zorko2011,Matsuda2012}, thus about $0.3-2$ K in the case of Fe$_{8}$, see Fig. \ref{S-mixing}. The oscillation periods of $\Delta_{m}$ vs $H_{x}$ depend on the magnitude and orientation of $\overrightarrow{d_{\rm{I,II}}}$: they increase with increasing $d_{\rm{I,II}}$ when $\overrightarrow{d_{\rm{I,II}}}$ points along $x$ but decrease when $\overrightarrow{d_{\rm{I,II}}}$ is along $y$. We have therefore set this vector along $x$ and varied $d_{\rm{I,II}}$ to fit the experimental susceptibility oscillations. As with the giant spin model, the dynamical susceptibility has been calculated by solving a Pauli master equation for the energy level populations of Eq.(\ref{hamilt2}). The best agreement, shown in Figs. \ref{XvsTHn0} and \ref{XvsH}, is found for $d_{\rm{I,II}}/k_{\rm B} = 1.28 \pm 0.05$ K, which is compatible with the above estimates. The model accounts for the oscillations observed in the ac susceptibility and related quantities ($T_{\rm b}$ and $\tau$) measured between $2$ K and $3$ K. In addition, it describes well the overall temperature dependence of $\Delta H_{x}$ between very low temperatures and $3$ K (Fig. \ref{XvsH}C). Finally, it predicts a ground state tunnel splitting $\Delta_{10}/k_{\rm B} = 10^{-7}$ K, which agrees with that determined from Landau-Zener experiments \cite{Wernsdorfer1999,Wernsdorfer2000}.

It can be concluded that Eq. (\ref{hamilt2}), despite its relative simplicity, agrees not only with available spectroscopic and magnetic data, but also provides a much more accurate description of the spin dynamics in Fe$_{8}$ than the giant spin model Eq. (\ref{hamilt1}). It also enables one to quantify the degree of $S$-mixing. The ground state of Eq.(\ref{hamilt2}) contains $88.4$ \% of $S=10$ states, $10.9$ \% of $S=9$ states and $0.7$ \% of states from other multiplets.


Summarizing, we have shown that the ac linear magnetic response and the superparamagnetic blocking of molecular nanomagnets are governed by quantum interferences, which can be tuned by an external magnetic field. Furthermore, the period of oscillations depends on the nature of the spin states involved in the tunneling processes, i.e. on whether they are pure ground-$S$ states or quantum superpositions with states from other multiplets. These results confirm that an accurate description of quantum phenomena in single molecule magnets should take into account quantum mixing between ground and excited multiplets. They also illustrate the sensitivity of interference phenomena to small changes in the wave function describing a physical system. By contrast, the relaxation time depends more strongly on the energy and number of such states. Ac susceptibility measurements performed under transverse magnetic fields provide then a rather powerful, and general, method for quantifying the degree of $S$-mixing in SMMs.


\begin{acknowledgments}
We acknowledge the assistance of Tomoaki Yamasaki and Dr. Miki Ueda in the synthesis of the samples. The present work has been partly funded trough the Spanish MINECO (grants MAT2012-38318-C03) and the Gobierno de Arag\'on (project MOLCHIP).
\end{acknowledgments}



\end{document}